\documentclass{appolb}
\usepackage{psfig}

\begin{document}
\title{Two pion electroproduction
\thanks{Presented at MESON'2000 Workshop, Krakow, May 19 - May 23, 2000}}
\author{ J.C. Nacher and E. Oset 
\address{Departamento de F\'{\i}sica Te\'orica and IFIC, 
Centro Mixto Universidad de Valencia-CSIC, Valencia, Spain}}

\maketitle
\vspace{-1cm}
\begin{abstract}
We have extended a model for the $\gamma N\rightarrow \pi\pi N$
reaction to virtual photons and selected the diagrams which have a $\Delta$
in the final state. 
The
agreement found with the $\gamma_v p\rightarrow\Delta^0\pi^+$ and  
$\gamma_v p\rightarrow\Delta^{++}\pi^-$ reactions is good. The sensitivity of 
the results to $N\Delta$ transition form factors is also studied. The present
reaction, selecting a particular final state, is an extra test for models of 
the 
$\gamma_{v} N\rightarrow \pi\pi N$ amplitude.
\end{abstract}

\section{Introduction}

The $\gamma N\rightarrow \pi\pi N$ reaction in nuclei has captured some
attention
recently and has proved to be a source of information on several aspects of 
resonance formation and decay as well as a test for chiral perturbation theory
at low energies. A model for the $\gamma p\rightarrow\pi^+\pi^- p$ reaction was
developed in [1] containing 67 Feynman diagrams by means of which a good 
reproduction of the cross section was found up to about $E_\gamma\simeq 1$ 
GeV.

A more reduced set of diagrams, with 20 terms , was found sufficient to describe
the reaction up to $E_\gamma\simeq 800$ MeV [2] where the Mainz 
experiments are done [3,4,5].

The 
extension of this kind of work to virtual photons should complement the 
knowledge obtained through the ($\gamma$,$2\pi$) and the related reactions. 
The coupling of the photons to the resonances depends on $q^2$ and
the dependence
can be different for different resonances. Hence, the interference of different 
mechanisms pointed above will depend on $q^2$ and with a sufficiently large 
range of $q^2$, one can pin down the mechanism of ($\gamma$,$2\pi$) with real 
or virtual photons with more precision than just with real photons, which would 
help settle the differences between present theoretical models.

However, there are already interesting two pion electroproduction
experiments selecting $\Delta$ in the final state. The
reactions are, $e p\rightarrow e^\prime\pi^-\Delta^{++}$
and $e p\rightarrow e^\prime\pi^+\Delta^0$ [6].
 It is thus quite interesting to extend present models of $(\gamma,2\pi)$
to the realm of virtual photons
and compare with existing data. In our paper [7] we do so, extending the model of ref.[2]
to deal with
the electroproduction process. This model is flexible enough and one can select
the diagrams which contain
$\Delta \pi$ in the final state in order to compare directly
with the measured cross sections.

The extension of the model requires three new ingredients: the introduction of the zeroth
component of the
photon coupling to resonances (calculations where done in [2] in the Coulomb gauge,
$\epsilon^0$, where the
zeroth component is not needed),  the implementation of the $q^2$
dependence of the amplitudes, which
will be discussed in forthcoming sessions, and the addition
of the explicit terms linked to the $S_{1/2}$ helicity amplitudes which
vanish for real photons.

Experiments on ($\gamma_v$,$2\pi$) are presently being done in the  Thomas 
Jefferson Laboratory [8], both for $N \Delta$ and $N\pi\pi$ production.

\section{Model for $e N\rightarrow e^\prime\Delta\pi$}
We will evaluate cross sections of virtual photons integrated over all the 
variables of the pions and the outgoing nucleon. In this case the formalism is 
identical to the one of inclusive $eN\rightarrow e^\prime X$
scattering [9,10]
or pion electroproduction after integrating over the pion variables 
[11,12]. 
For the model of the $\gamma_v N \rightarrow\Delta\pi$ reaction we take the 
same diagrammatic approach as in ref.[2] and select the diagrams which have a $\Delta$ in the 
final state. The diagrams which  contribute to the process are depicted in 
fig.1

 \begin{figure}[h]
\centerline{\protect
\hbox{
\psfig{file=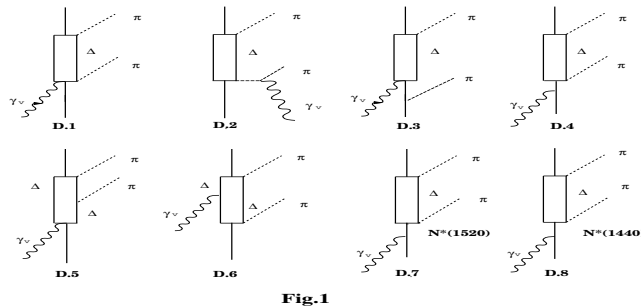,height=4cm,width=8.5cm,angle=0}}}
\caption{Feynman diagrams used in the model for $\gamma_v p\rightarrow
\pi\Delta$}
\end{figure}

We follow the paper from Devenish et al. [13] in our approach to electromagnetic
transitions for Roper and $N^\ast(1520)$ resonances.
As we are working with virtual photons we need to care about these
couplings and hence include terms which vanish for real photons.

Gauge invariance is one of the important elements 
in a model involving photons and implies that
\begin{equation}
T^\mu q_\mu = 0
\end{equation}

 However, as discussed in the study of the $e 
N\rightarrow e^\prime N\pi$ reaction in [14], and as can be easily seen by inspection of the diagrams
and the amplitudes, the constraint of eq. (1) still requires
the equality of four electromagnetic form factors,

\begin{equation}  
F_1^p(q^2) = F_1^\Delta(q^2)= F_{\gamma\pi\pi} = F_c(q^2)
\end{equation}

The form factors of eq. (2) are respectively
 the $\gamma N N$, $\gamma\Delta\Delta$, $\gamma\pi\pi$ and 
 $\gamma\Delta N\pi$ ones.
 These form factors are usually parametrized in different forms, except for
$F_1^p (q^2)$ and $F_1^\Delta (q^2)$ which are taken equal, 
as it would come from
ordinary quark models.

Although the model is gauge invariant with the prescription
of eq. (2) there is the inconvenience that the results
depend upon which one of the three form factors we take for all of them.

 We should note however, that the dominant term, by large, is the 
$\Delta$ Kroll Ruderman 
and pion pole terms. This is also so in the test of gauge invariance 
where the two terms involving 
the $F_1^p(q^2)$ form factor in diagrams D4, D6 give only recoil 
contributions of the order of 
O($p_\pi$/m) in eq. (1). This justifies the use of $F_c(q^2)$ or $F_{\gamma\pi\pi}(q^2)$ for all the 
form factors.

There is, however, another way to respect gauge invariance, while at the same
time using different form factors which is proposed in \cite{berends} and 
to which we refer in what follows as Berends et al. approach. 

\section{Results and conclusions}
We have tested our results [7] with the experimental data of refs. [6,11]. We show
the cross section of $\gamma_v p\rightarrow\Delta^{++}\pi^-$ and
$\gamma_v p\rightarrow\Delta^0\pi^+$ ($\Delta^0\rightarrow\pi^- p$), as a function
of W, the virtual photon-proton ($\gamma_v p$) center of mass energy, and for
different values of $Q^2$.
We have made different calculations. One of them corresponds
to using all form factors equal (which we set to $F_{\gamma\pi\pi}$)
with two different values of $\lambda_\pi^2$, 0.5 $GeV^2$ and 0.6 $GeV^2$. 
In [7] we see that the cross section increases by about 10 $\%$ when going from
$\lambda_\pi^2$=0.5 $GeV^2$ and $\lambda_\pi^2$=0.6 $GeV^2$. We also
show the results taking $F_1^p$, $F_1^\Delta$ and setting $F_c=F_{\gamma\pi\pi}$ with $\lambda_\pi^2$=0.6
 $GeV^2$. This latter calculation is not gauge invariant. However we see 
that the deviation with respect to the gauge invariant one assuming all form
factors equal is very small [7]. This reflects the fact that the relevant terms
in the model are those involving $F_{\gamma\pi\pi}$ and $F_c$, the pion pole and
$\Delta$ Kroll Ruderman terms.

 We also evaluate the cross section using Berends gauge invariant approach
with different form factors \cite{berends}. We show the results in fig. 2.
The continuous line in the figure
is obtained with this prescription using $F_1^p$, $F_1^\Delta$ but setting $F_c=F_{\gamma\pi\pi}$ with
$\lambda_\pi^2$ = 0.5 $GeV^2$.

We see that these results are remarkably similar to those 
where $F_c$ and $F_{\gamma\pi\pi}$ had the same
values as here but $F_1^p$, $F_1^\Delta$ were set equal to
$F_{\gamma\pi\pi}$ in order to preserve gauge invariance.

The dotted line in fig. 2 corresponds to the same parametrization for
$F_c$ as for $F_{\gamma\pi\pi}$ but parameter $\lambda_c^2$= 0.8 $GeV^2$.
This shows the sensitivity of the results to $F_c$  which appears in the 
dominant Kroll-Ruderman term.

\begin{figure}[h]
\centerline{\protect
\hbox{
\psfig{file=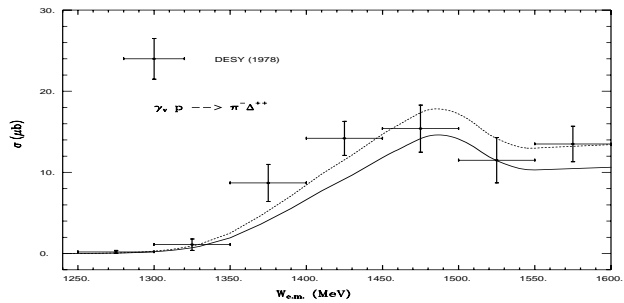,height=4cm,width=8.5cm,angle=-90}}}
\caption{\small{Cross sections from $\gamma_v p\rightarrow\Delta^{++}\pi^-$ with
Berends formalism included.
Continuous line: $F_1^p, F_1^\Delta, F_{\gamma\pi\pi}=F_c$ with
$\lambda_\pi^2= 0.5$ $GeV^2$. Dotted line: same as continuous line with the
parameter for $F_{\gamma\pi\pi}$ 0.5 $GeV^2$ and for $F_c$ is 0.8 $GeV^2$.}}
\end{figure}
 In summary we could remark the following points:
We have shown in [7] that the peak in the cross section is due to an interference
between the $\Delta$ Kroll Ruderman term and the $N^\ast(1520)$ excitation
process followed by $\Delta\pi$ decay.
Different sets of form factors have been used in our model in order to show the
sensitivity of the results to these changes. These tests should
be useful in view of the coming data and the possibility to extract relevant 
information from them. We have calculated the 
separation
of the transverse and longitudinal cross sections and found that the
transverse one largely dominates the cross sections.
Finally, it is also interesting to note that the present model is just part of a more general 
$\gamma_v N \rightarrow\pi\pi N$ model which selects only the terms 
where a $\pi N$ pair of the final state appears
forming a $\Delta$ state.

\end{document}